\documentstyle[preprint,aps,floats,epsfig]{revtex}

\begin{document}

\draft

\preprint{\rightline{ANL-HEP-PR-01-028 \hspace{1cm} SWAT/01/306}}

\title{Two-colour QCD at non-zero quark-number density.}

\author{J.~B.~Kogut}
\address{Dept. of Physics, University of Illinois, 1110 West Green Street,
Urbana, IL 61801-3080, USA}
\author{D.~K.~Sinclair}
\address{HEP Division, Argonne National Laboratory, 9700 South Cass Avenue,
Argonne, IL 60439, USA}
\author{S.~J.~Hands and S.~E.~Morrison}
\address{Department of Physics, University of Wales 
        Swansea, Singleton Park, Swansea SA2 8PP, UK}

\maketitle

\begin{abstract}
We have simulated two-colour four-flavour QCD at non-zero chemical potential
$\mu$ for quark number. Simulations were performed on $8^4$ and $12^3 \times
24$ lattices. Clear evidence was seen for the formation of a colourless
diquark condensate which breaks quark number spontaneously, for $\mu > \mu_c
\sim m_\pi/2$. The transition appears to be second order. We have measured the
spectrum of scalar and pseudoscalar bosons which shows clear evidence for the
expected Goldstone boson. Our results are in qualitative agreement with those
from effective Lagrangians for the potential Goldstone excitations of this
theory.
\end{abstract}

\newpage

\section{Introduction}

QCD at finite quark/baryon-number density at zero and at finite temperature
describes nuclear matter. Nuclear matter at high temperatures (and possibly
densities) was certainly present in the early universe. Neutron stars consist
of dense cold nuclear matter. RHIC and the CERN heavy-ion program promise to
produce hot nuclear matter in the laboratory. Calculating the properties of
high density nuclear matter could predict if and where strange matter could be
produced. Any method which can be used to determine the properties of nuclear
matter could be adapted to nuclear physics calculations.

Finite quark-number density is best achieved by introducing a chemical 
potential $\mu$ for quark-number, and using the grand-canonical partition
function. Unfortunately, this renders the Euclidean-time fermion determinant
complex, with a real part which can change sign. Standard lattice simulations,
which rely on importance sampling, fail in this case. Attempting to circumvent
these problems by using canonical (fixed quark number) ensembles fail except
at high temperatures \cite{kekl} because of sign problems.

Until a simulation method is found which avoids these difficulties, it is
useful to study models which exhibit {\it some} of the properties of QCD at
high $\mu$. Now it is expected that, at zero temperature, nuclear matter
undergoes a phase transition at $\mu$ of order one third the mass of the
nucleon. It has been proposed that at still higher $\mu$ the ground state
is characterised by a diquark condensate
\cite{b,bl,arw,rssv,as}.
Such a condensate would not only cause spontaneous
breaking of baryon number, but would also spontaneously break colour. Since
colour is a gauge symmetry, such breaking is realized in the Higgs mode. Thus
nuclear matter would become a colour superconductor at high $\mu$. 

For this reason we have simulated 2-colour QCD, i.e. SU(2) Yang-Mills theory
with fermion matter fields (`quarks') in the fundamental representation of
$SU(2)_{colour}$, and finite $\mu$. As well as having colour confinement this
theory does exhibit diquark condensation as we shall demonstrate in this
paper, but for $\mu> \mu_c \sim m_\pi/2$, since the diquark `baryons' in the
same multiplet as the pions, also have mass $m_\pi$. For $\mu \lesssim m_\pi$
the phenomenon 
is describable as a rotation of the condensate from the chiral to the
diquark direction as predicted by effective Lagrangian analyses \cite{kstvz}.
Unlike in true (3-colour) QCD, the diquark condensates are colourless, and the
broken symmetry is realized in the Goldstone mode, and there is no colour
superconductivity, but rather superfluidity, as in liquid $^3$He. 
Despite this, we shall later argue that this theory is more
similar to 2-flavour QCD than one might think (see section 4).

Since 2-colour QCD has a non-negative determinant and pfaffian, even at
non-zero $\mu$, standard simulation methods can be used. We use the hybrid
molelcular-dynamics method and simulate the theory with 4 flavours of staggered
quarks \cite{hmd}. Pfaffian simulations of a 4-fermion model including a
diquark source term have been reported in \cite{hlm}.
We have run simulations at a moderately large quark mass and an
intermediate gauge coupling on $8^4$ and $12^3 \times 24$ lattices, i.e. at
zero temperature. We measured order parameters including the chiral and
diquark condensates, the quark-number and energy densities, and the Wilson
Line (Polyakov loop). The larger lattice allowed us to observe finite size
effects and, more importantly to measure the scalar and pseudoscalar meson and
diquark masses, which include all the potential Goldstone bosons in the
theory. Preliminary results from these simulations were reported at
Lattice'2000, Bangalore \cite{hkms}. 
The extension of these calculations to finite temperature
was reported in a recent letter \cite{kts}.
This work builds on early work with 8 quark flavours which presented far less
conclusive results \cite{hklm}. Previous studies of diquark condensation in
this model with various numbers of flavours have either used the approximation
where $\lambda=0$ in the updating algorithm \cite{l=0} (as does \cite{hklm}),
or been in the strong gauge-coupling regime \cite{sg}.

Section 2 introduces the staggered-fermion lattice port of 2-colour QCD. The
results of our simulation are presented in section 3. Section 4 gives our
conclusions and indicates future avenues of research.

\section{Lattice 2-colour QCD}

Because in 2-colour QCD fundamental quarks and antiquarks lie in the same
representation of $SU(2)_{colour}$ the flavour symmetry group for $N_f$
flavours is enlarged from $SU_L(N_f) \times SU_R(N_f) \times U_V(1)$ to
$SU(2N_f)$. The pattern of chiral symmetry breaking is 
$SU(2N_f) \rightarrow Sp(2N_f)$ rather than the usual 
$SU_L(N_f) \times SU_R(N_f) \rightarrow SU_V(N_f)$ \cite{pg,kstvz}. 
3-colour QCD with 1 staggered quark (4 continuum flavours) has a $U_V(1)
\times U(1)_\varepsilon$ flavour symmetry. For 2 colours this is enhanced to 
$U(2)$ \cite{hklm}. 
Chiral or quark-number symmetry
breaking (the chiral and diquark condensates lie in the same U(2) multiplet)
occurs according to the pattern $U(2) \rightarrow U(1)$.

The quark action for 2-colour QCD with one staggered quark is
\begin{equation}
S_f = \sum_{sites}\left\{\bar{\chi}[D\!\!\!\!/\,(\mu) + m]\chi 
+ \frac{1}{2}\lambda[\chi^T\tau_2\chi + \bar{\chi}\tau_2\bar{\chi}^T]\right\}
\label{eqn:lagrangian}
\end{equation} 
where $D\!\!\!\!/\,(\mu)$ is the normal staggered covariant finite difference
operator with $\mu$ introduced by multiplying the
links in the $+t$ direction
by $e^\mu$ and those in the $-t$ direction by $e^{-\mu}$. The superscript $T$
stands for transposition. Note that we have introduced a gauge-invariant
Majorana mass $\lambda$ which explicitly breaks quark-number symmetry. Such an
explicit symmetry breaking term is needed to observe spontaneous symmetry
breaking on a finite lattice. We shall be interested in the limit 
$\lambda \rightarrow 0$. Integrating out these fermion fields yields
\begin{equation}
pfaffian\left[\begin{array}{cc} \lambda\tau_2     &    {\cal A}  \nonumber   \\
                               -{\cal A}^T        &    \lambda\tau_2
\end{array}\right] = \sqrt{{\rm det}({\cal A}^\dagger {\cal A} + \lambda^2)}
\label{eqn:pfaffian}
\end{equation}
where
\begin{equation}
           {\cal A} \equiv  D\!\!\!\!/\,(\mu)+m
\end{equation}
We note that ${\cal A}^\dagger {\cal A} + \lambda^2$ is positive definite for
finite $\lambda$. Hence the pfaffian never vanishes and thus, by continuity
arguments never changes sign and can be chosen to be positive. Note that
${\rm det}({\cal A})$ has been shown to be positive in \cite{hmmoss}.
Denoting the
$2 \times 2$ matrix in equation~\ref{eqn:pfaffian} by ${\cal M}$ we have seen
that its determinant is the determinant of a positive definite matrix and thus
can be used directly in our simulations. To do this we define
$\widetilde{\cal M}$ by
\begin{equation}
\widetilde{\cal M}  = \left[\begin{array}{cc}    1       &      0            \\
                                                 0       &   \tau_2
                                                             \end{array}\right]
          {\cal M}    \left[\begin{array}{cc} \tau_2     &      0            \\ 
     
                                                 0       &      1 
                                                             \end{array}\right]
\end{equation}
so that
\begin{equation}
\widetilde{\cal M} = \left[\begin{array}{cc} \lambda     &    {\cal A}       \\
                                      -{\cal A}^\dagger  &     \lambda
                                                             \end{array}\right]
\end{equation}
and
\begin{equation}
\widetilde{\cal M}^\dagger\widetilde{\cal M} \rightarrow \left[\begin{array}{cc}
{\cal A}^\dagger {\cal A} + \lambda^2  &                   0                  \\
                 0                     &  {\cal A} {\cal A}^\dagger + \lambda^2
                                            \end{array}\right].
\end{equation}
We note that $\det[{\cal A} {\cal A}^\dagger + \lambda^2] =
\det[{\cal A}^\dagger {\cal A} + \lambda^2]$. Because we are now dealing
with the matrix $\widetilde{\cal M}^\dagger\widetilde{\cal M}$
we can use the hybrid molecular dynamics method with `noisy' fermions \cite{hmd}
to simulate this theory. Here, although we generate gaussian noise with both
upper and lower components, we keep only the upper components of the
pseudo-fermion field to calculate $\det[{\cal A}^\dagger {\cal A} + \lambda^2]$,
which means that we only need to invert ${\cal A}^\dagger {\cal A} + \lambda^2$
at each update. Keeping only half the components of the pseudo-fermion field
is entirely analogous to keeping only the fermion fields on even sites in normal
QCD simulations. The square root of equation~\ref{eqn:pfaffian} is obtained by
inserting a factor of $\frac{1}{2}$ in front of the fermion term in the 
stochastic action in the standard manner.

We now give a brief discussion of symmetry breaking in this model. This is
covered in more detail in \cite{hklm}.
At $\mu=m=\lambda=0$, the $U(2)$ symmetry will break spontaneously.
Two directions in which it will choose to break are of particular interest.
The first is where it breaks to give a non-zero chiral condensate
$\langle\bar{\chi}\chi\rangle$. There will be 3 Goldstone bosons corresponding
to the 3 broken generators of $U(2)$. These
states and their corresponding $U(2)$
generators are
\begin{eqnarray}
{\bf 1}  &\Longrightarrow& \bar{\chi}\epsilon\chi     \nonumber            \\
\sigma_1 &\Longrightarrow& \chi^T\tau_2\chi - \bar{\chi}\tau_2\bar{\chi}^T \\
\sigma_2 &\Longrightarrow& \chi^T\tau_2\chi + \bar{\chi}\tau_2\bar{\chi}^T
\nonumber
\end{eqnarray}
$\sigma_3$ remains unbroken. The second is where $U(2)$ breaks to give a 
non-zero diquark condensate 
$\frac{1}{2}\langle\chi^T\tau_2\chi + \bar{\chi}\tau_2\bar{\chi}^T\rangle$.
This time the 3 Goldstone bosons and corresponding generators are
\begin{eqnarray}
{\bf 1}  &\Longrightarrow& \chi^T\tau_2\epsilon\chi + 
                           \bar{\chi}\tau_2\epsilon\bar{\chi}^T \nonumber \\
\sigma_2 &\Longrightarrow& \bar{\chi}\chi                                 \\
\sigma_3 &\Longrightarrow& \chi^T\tau_2\chi - \bar{\chi}\tau_2\bar{\chi}^T
\nonumber
\end{eqnarray}
$\sigma_1$ remains unbroken. When $\mu \ne 0$ only 2 Goldstone bosons remain,
$\chi^T\tau_2\epsilon\chi + \bar{\chi}\tau_2\epsilon\bar{\chi}^T$ and
$\chi^T\tau_2\chi - \bar{\chi}\tau_2\bar{\chi}^T$. When in addition 
$m \ne 0$, only the latter state remains a Goldstone boson. A more detailed
study of the patterns of symmetry breaking can be performed in terms of an
effective Lagrangian for the Goldstone modes as in \cite{kstvz}. The only 
difference is that the effective field, denoted $\Sigma$ in that work, here
belongs to a symmetric $2 \times 2$ tensor representation of $U(2)$ rather
than to the antisymmetric $2N_f \times 2N_f$ representation of $SU(2N_f)$ of
the continuum case.

One can see the remnant $U(1)$ symmetry when $\lambda \rightarrow 0$ by
allowing $\lambda$ to become complex. The Majorana mass term in 
equation~\ref{eqn:lagrangian} then becomes
$\frac{1}{2}[\lambda\chi^T\tau_2\chi + \lambda^*\bar{\chi}\tau_2\bar{\chi}^T]$.
$\lambda^2$ is replaced by $|\lambda|^2$ in the pfaffian.

Although the 2-flavour theory would be of more interest in the continuum, we
have chosen to simulate the 4-flavour theory because this represents a single
staggered quark species and thus has well defined symmetries and a well defined
spectrum at all lattice spacings. Unlike the 8-flavour case simulated in
\cite{hklm} it probably does have a sensible continuum limit.

\section{Simulations of 2-colour, 4-flavour lattice QCD}

We have simulated 2-colour QCD with 1 staggered quark species (4 continuum
flavours) on $8^4$ and $12^3 \times 24$ lattices. The simulations reported
here are all at $\beta=4/g^2=1.5$ which is roughly the $\beta_c$ for the
finite temperature transition on an $N_t=4$ lattice \cite{kogut}. 
This first set of
simulations has been performed with quark mass $m=0.1$ in lattice units.
These simulations are currently being repeated at $m=0.025$, where the smaller
pion mass will give a richer spectrum of Goldstone and pseudo-Goldstone bosons
and where a larger portion of the relevant phase diagram should be described
by effective Lagrangians. (We have also performed some zero temperature 
simulations at $\beta=1.0$ and $m=0.05$. This has been reported in our finite
temperature/finite $\mu$ letter \cite{kts}.) 
Since we wished to take the limit 
$\lambda \rightarrow 0$, we needed $\lambda << m$. The values we chose were
$\lambda=0.01$ and $\lambda=0.02$ for $m=0.1$. (At low $\mu$'s we also ran at
$\lambda=0$.) 

\begin{figure}[htb]
\epsfxsize=4in
\centerline{\epsffile{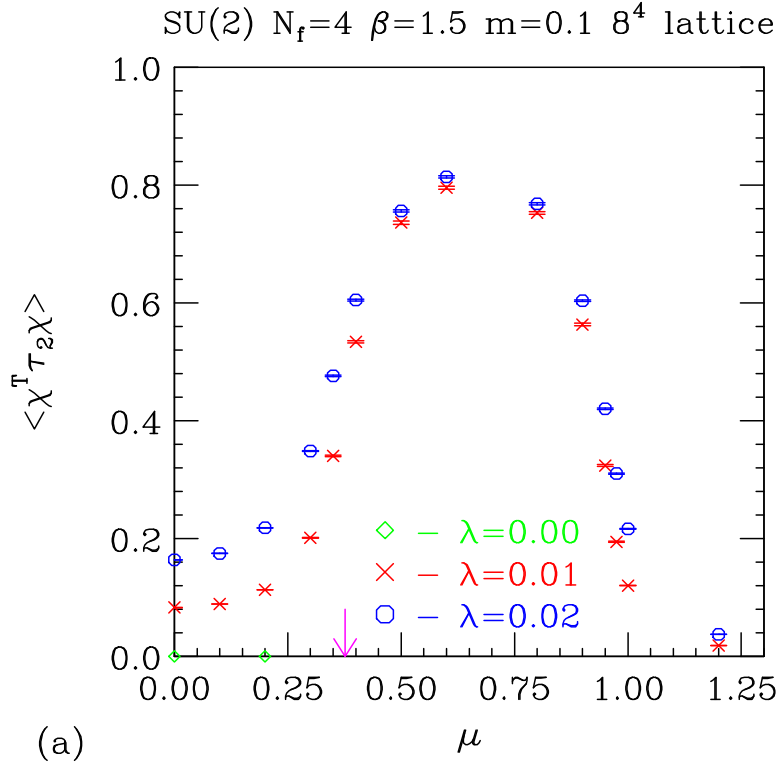}}
\vspace{0.25in}
\centerline{\epsffile{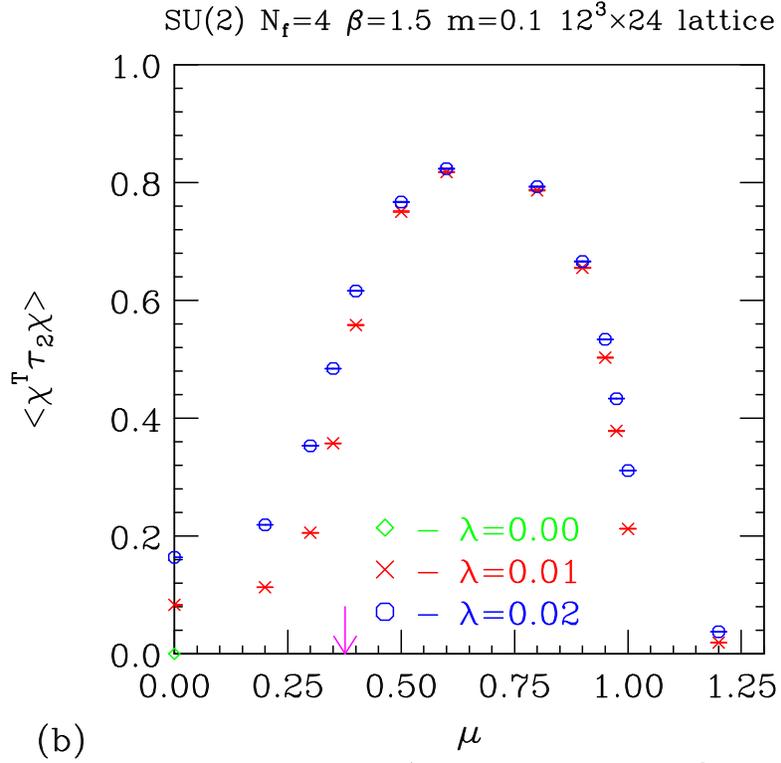}}
\caption{$\langle\chi^T\tau_2\chi\rangle$ as a function of $\mu$:
a) on an $8^4$ lattice and b) on a $12^3 \times 24$ lattice. The arrow is at
$\mu=m_\pi$}%
\label{fig:pt2p}
\end{figure}

The smaller lattice was used to map out the interesting range of
$\mu$ values, measuring order parameters including the diquark condensate
$\langle\chi^T\tau_2\chi\rangle$, the chiral condensate
$\langle\bar{\chi}\chi\rangle=\langle\bar{\psi}\psi\rangle$, and the number
density $j_0=\frac{1}{V}{\partial \ln S_f \over \partial\mu}$. In addition to
measuring these quantities on the larger lattice, we also calculated the
spectrum of potential Goldstone bosons. The length of each `run' was 2000
molecular dynamics time units. $dt$ had to be chosen as low as $0.005$ for
$\lambda=0.01$ and $0.4 < \mu \le 0.975$. In figure~\ref{fig:pt2p} we have 
plotted the diquark condensate as a function of $\mu$ at each $\lambda$ for both
lattice sizes.

\begin{figure}[htb]                                                             
\epsfxsize=7in                                                                  
\centerline{\epsffile{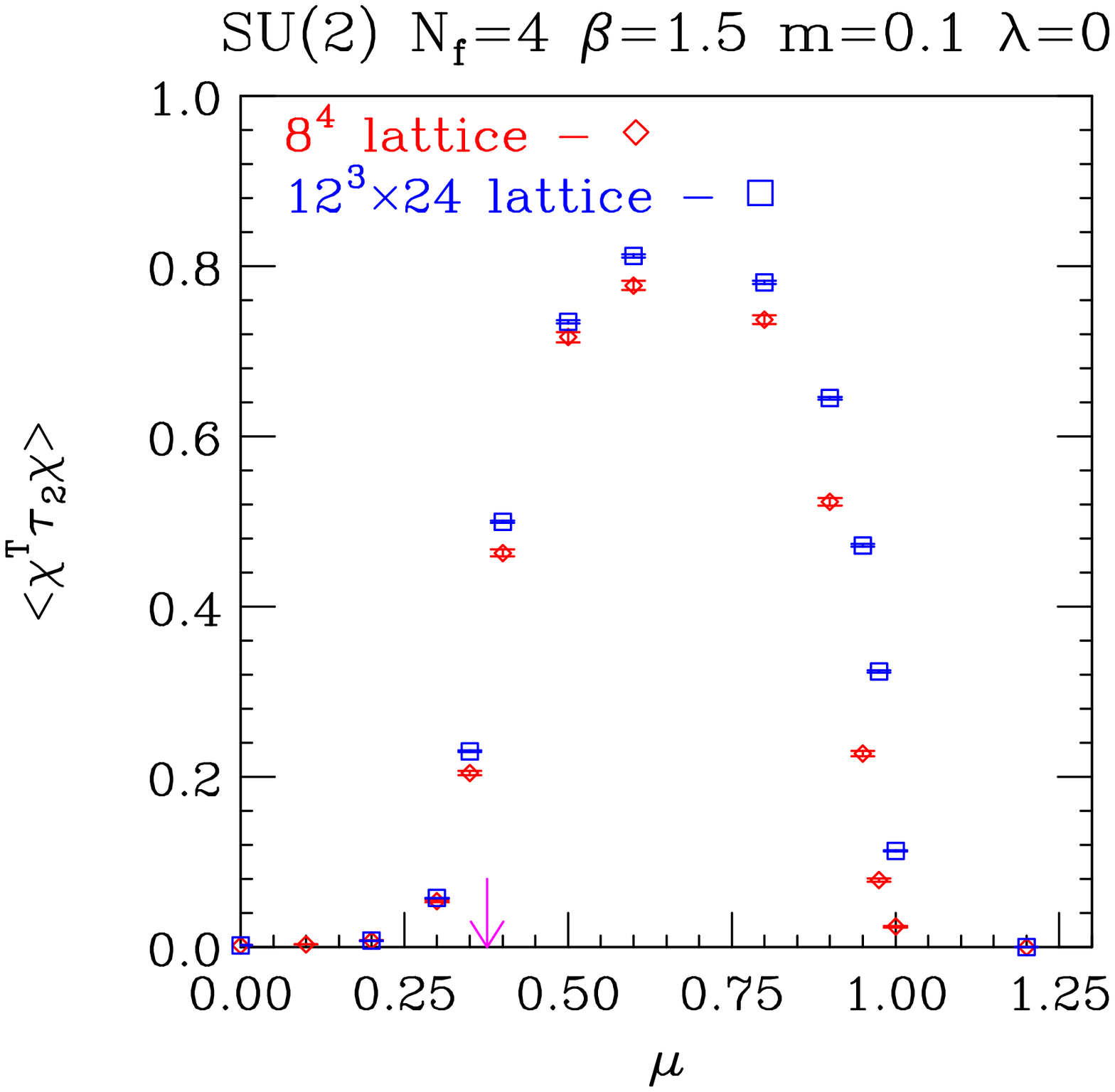}}                            
\caption{$\langle\chi^T\tau_2\chi\rangle$ extrapolated to $\lambda=0$ as a
function of $\mu$ on both $8^4$ and $12^3 \times 24$ lattices.}%
\label{fig:pt2p_0}
\end{figure}

Since we are interested in the limit where the symmetry breaking parameter
$\lambda \rightarrow 0$, we have performed a linear extrapolation of the
diquark condensate to $\lambda=0$. Note that the effective Lagrangian 
calculations \cite{kstvz} suggest that linear extrapolations are the correct
approach for $\lambda$ sufficiently small, except at $\mu_c$. The results of
these extrapolations are plotted in figure~\ref{fig:pt2p_0}.
What we first notice is that for $\mu \le 0.2$, the extrapolated diquark
condensate is small enough that we can believe that it should be zero. For
$\mu \ge 0.35$ it is clearly non-zero. The points at $\mu=0.3$ would appear
to show finite size rounding were it not for the fact that the $8^4$ and
$12^3 \times 24$ points are so close together. We think it more likely that
$\mu=0.3$ is so close to the transition that the linear extrapolation has
broken down. We therfore conclude that the system undergoes a phase transition
at $\mu = \mu_c \approx 0.3 < m_\pi/2 = 0.37622(5)$. If this holds true, the
fact that $\mu_c$ is less than $m_\pi/2$ would indicate that the diquark
`baryons' do not exist as free particles but bind into `nuclear' matter. Over
the range $0.35\le\mu\le0.6$, the condensate increases with a curvature
consistent with a critical index $\beta_{qq} < 1$ (the tree level effective
Lagrangian analysis predicts the mean field result $\beta_{qq} =
\frac{1}{2}$). Since the condensate starts to decrease soon after $\mu=0.6$,
the scaling region is narrow and it would require more points to even try to
extract this critical index. For $\mu \gtrsim m_\pi$, the condensate starts to
decrease, approaching zero for large $\mu$, which would appear to be a
saturation effect. In fact figure~\ref{fig:pt2p_0} suggests that the condensate
vanishes for $\mu > \mu_s \approx 1$, indicating a saturation phase transition.
We shall have more to say about this later.
Note that this decrease in the condensate with $\mu$ is
not predicted by the effective Lagrangian analysis, which is only expected to
be valid for small $\mu$ and $m_\pi$. This is not surprising if it is indeed a
saturation effect. Saturation is a result of the fermi statistics of the
quarks, and should not be seen in a model which only considers the system's
bosonic excitations.

\begin{figure}[htb]
\epsfxsize=4in
\centerline{\epsffile{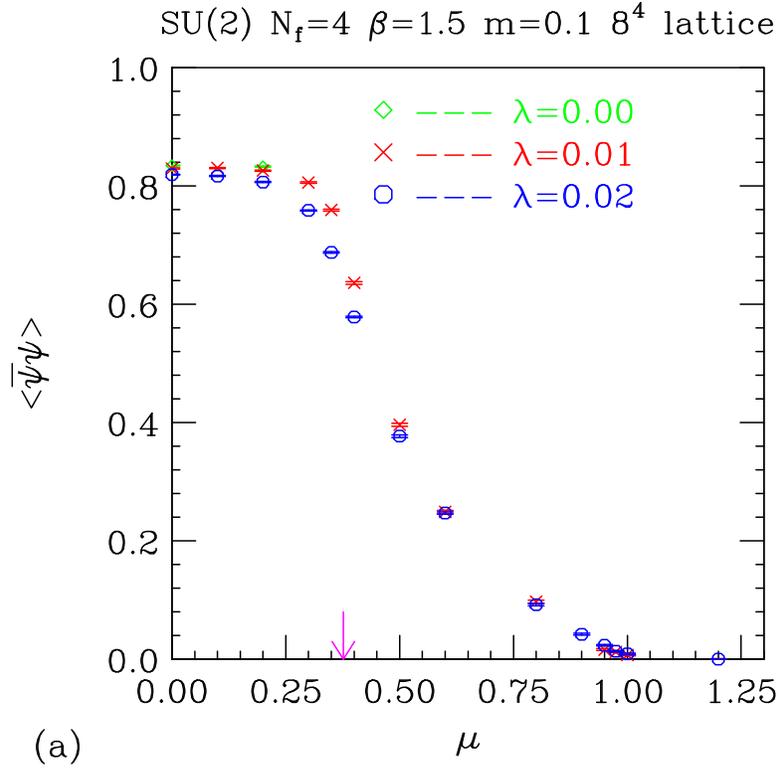}}
\vspace{0.25in}
\centerline{\epsffile{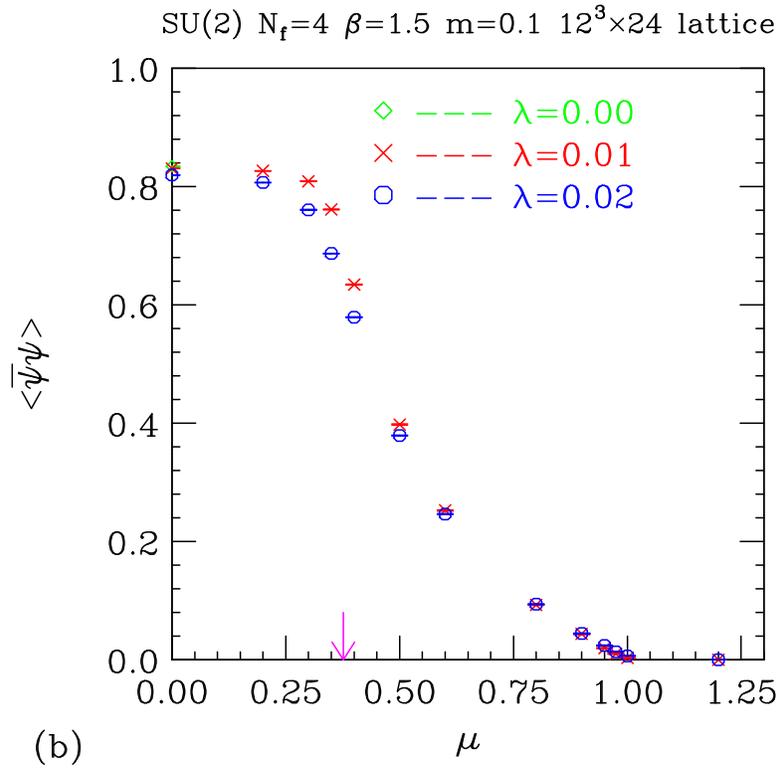}}
\caption{$\langle\bar{\psi}\psi\rangle$ as a function of $\mu$:
a) on an $8^4$ lattice and b) on a $12^3 \times 24$ lattice.}%
\label{fig:pbp}
\end{figure}

We now turn to a consideration of the chiral condensate,
$\langle\bar{\chi}\chi\rangle=\langle\bar{\psi}\psi\rangle$. This is plotted in
figure~\ref{fig:pbp} for our 2 different lattice sizes. 
In the $\lambda \rightarrow 0$ limit, this is expected to be constant at its
$\mu=0$ value for $\mu < \mu_c$. These plots are consistent with this
expectation. Above $\mu_c$, effective Lagrangian studies predict that the
condensate merely rotates from the chiral direction to the diquark direction,
so that the magnitude of the condensate, i.e.
$\sqrt{\langle\bar{\chi}\chi\rangle^2+\langle\chi^T\tau_2\chi\rangle^2}$,
should remain constant and independent of $\lambda$. Since the diquark
condensate increases up to $\mu \sim m_\pi$, this means that the chiral
condensate should fall, as it does. It, however, continues to fall past this
point, because of saturation effects, appearing to vanish for $\mu > \mu_s$.
To test the prediction that the principal effect for $\mu < m_\pi$ is merely a
rotation, we have tabulated
$\sqrt{\langle\bar{\chi}\chi\rangle^2+\langle\chi^T\tau_2\chi\rangle^2}$ in
table~\ref{tab:condensate}.
\begin{table}[htb]
\begin{tabular}{ccc}
        &   \multicolumn{2}{c}{$|{\rm condensate}|$}    \\
$\mu$   &   $\lambda=0.01$ &  $\lambda=0.02$    \\
\hline
0.000   &   0.835          &  0.836             \\
0.200   &   0.834          &  0.836             \\
0.300   &   0.835          &  0.839             \\
0.350   &   0.841          &  0.840             \\
0.400   &   0.845          &  0.846             \\
0.500   &   0.850          &  0.856             \\
0.600   &   0.856          &  0.860             \\
0.800   &   0.792          &  0.799             \\
0.900   &   0.657          &  0.667             \\
0.950   &   0.503          &  0.534             \\
0.975   &   0.379          &  0.433             \\
1.000   &   0.212          &  0.311             \\
1.200   &   0.019          &  0.038
\end{tabular}
\caption{Magnitude of the condensate
$\sqrt{\langle\bar{\chi}\chi\rangle^2+\langle\chi^T\tau_2\chi\rangle^2}$
as functions of $\mu$ and $\lambda$ on a $12^3 \times 24$ lattice. Errors are
not quoted but they are in the next or a subsequent digit after the least 
significant quoted digit.}
\label{tab:condensate}
\end{table}
We see that, for $\mu \le 0.6$, the magnitude of the condensate is approximately
constant and independent of $\lambda$, so that the main effect is a rotation of
the condensate from the direction of chiral symmetry breaking to that of
quark-number breaking. For $\mu \ge 0.8$, it decreases with increasing $\mu$
approaching zero for large $\mu$.

\begin{figure}[htb]
\epsfxsize=4in
\centerline{\epsffile{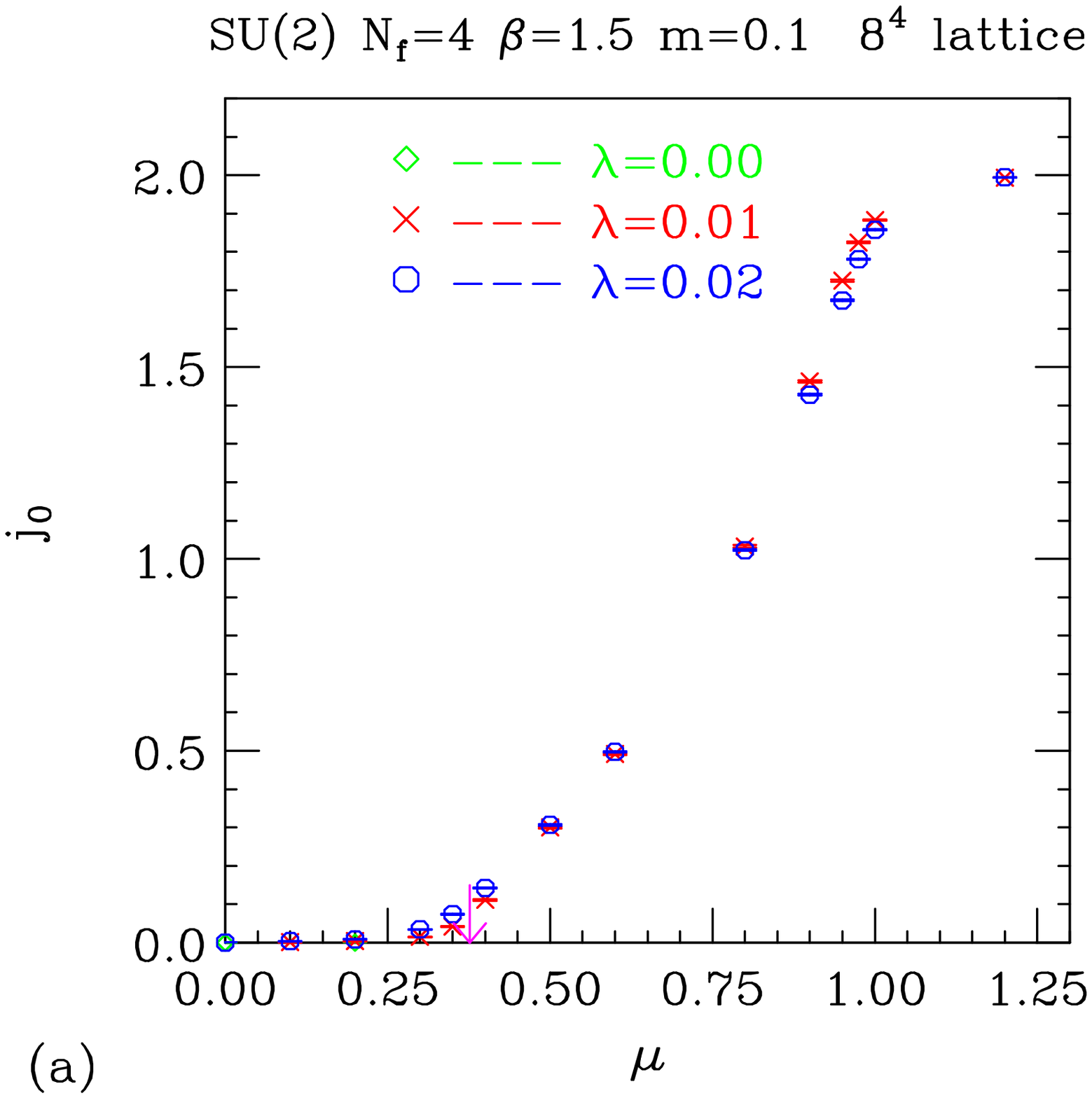}}
\vspace{0.25in}
\centerline{\epsffile{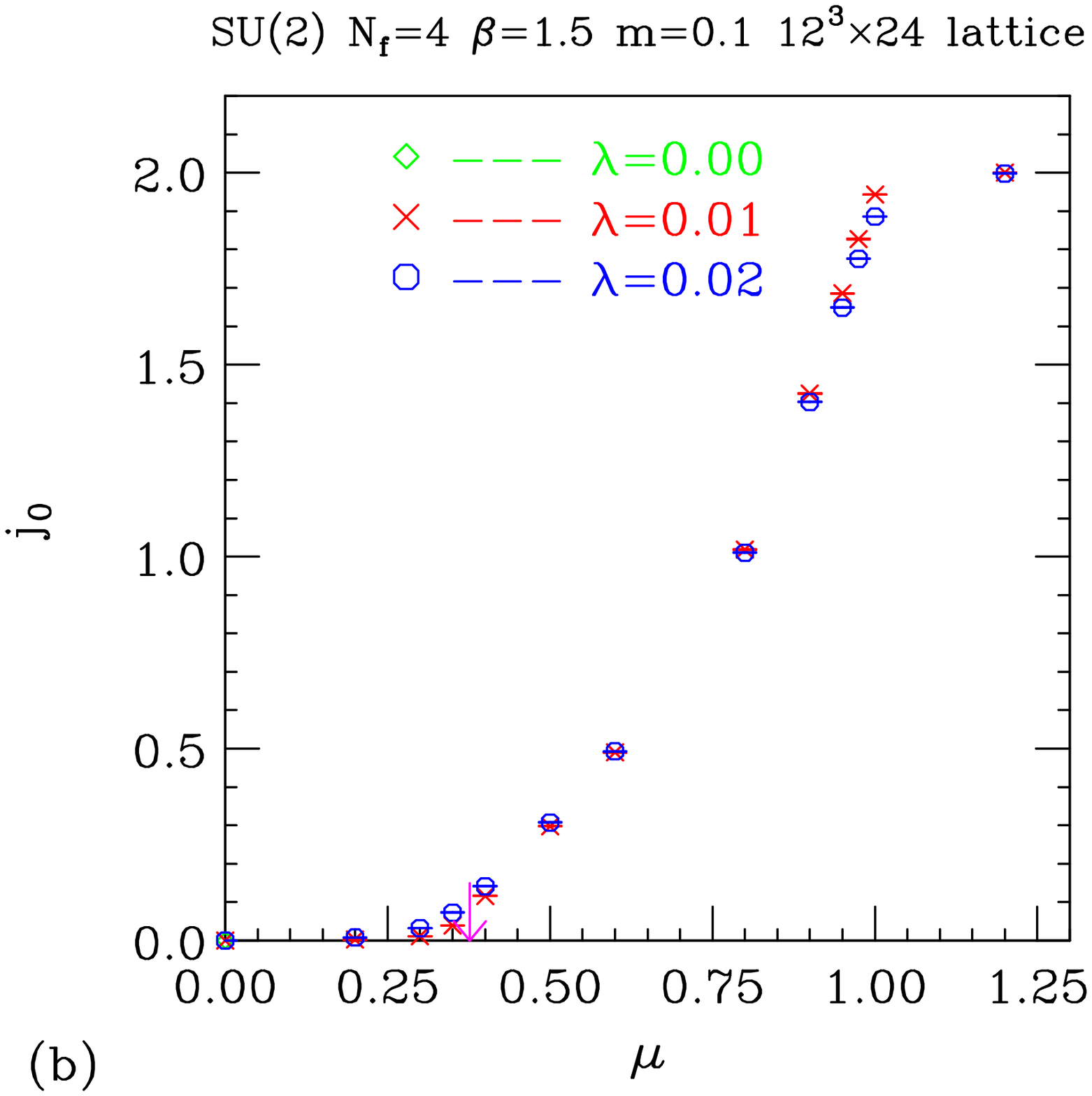}}
\caption{Quark-number density as a function of $\mu$:
a) on an $8^4$ lattice and b) on a $12^3 \times 24$ lattice.}%
\label{fig:j0}
\end{figure}

In figure~\ref{fig:j0} we plot the quark-number density 
$j_0=\frac{1}{V}{\partial \ln S_f \over \partial\mu}$ as a function of $\mu$ for
our 2 lattices.
$j_0$ is consistent with zero as $\lambda \rightarrow 0$, for $\mu < \mu_c$.
(Note effective Lagrangians predict that it vanishes quadratically with
$\lambda$ in this region.) Above $\mu_c$ it increases with increasing $\mu$
approaching the saturation value of 2 (1 staggered quark field of each
colour/site, the maximum value allowed by fermi statistics) for large $\mu$.
Note that figure \ref{fig:j0} suggests that $j_0=2$ for $\mu > \mu_s$ with
$\mu_s$ consistent with that predicted from the condensates. This indicates
that it is this saturation which causes the condensates to vanish for
$\mu > \mu_s$. It also tells us that this transition is an artifact of the
finite lattice spacing and would recede to infinite $\mu$ in the continuum
limit.

On the $12^3 \times 24$ lattice we have measured the propagators for all the
potential Goldstone bosons, both scalar and pseudoscalar using a single noisy
point source for the connected propagators and multiple (5) noisy point
sources for the disconnected propagators. We have measured both diagonal and
off-diagonal zero momentum propagators.

\begin{figure}[htb]                                                          
\epsfxsize=7in                                                                  
\centerline{\epsffile{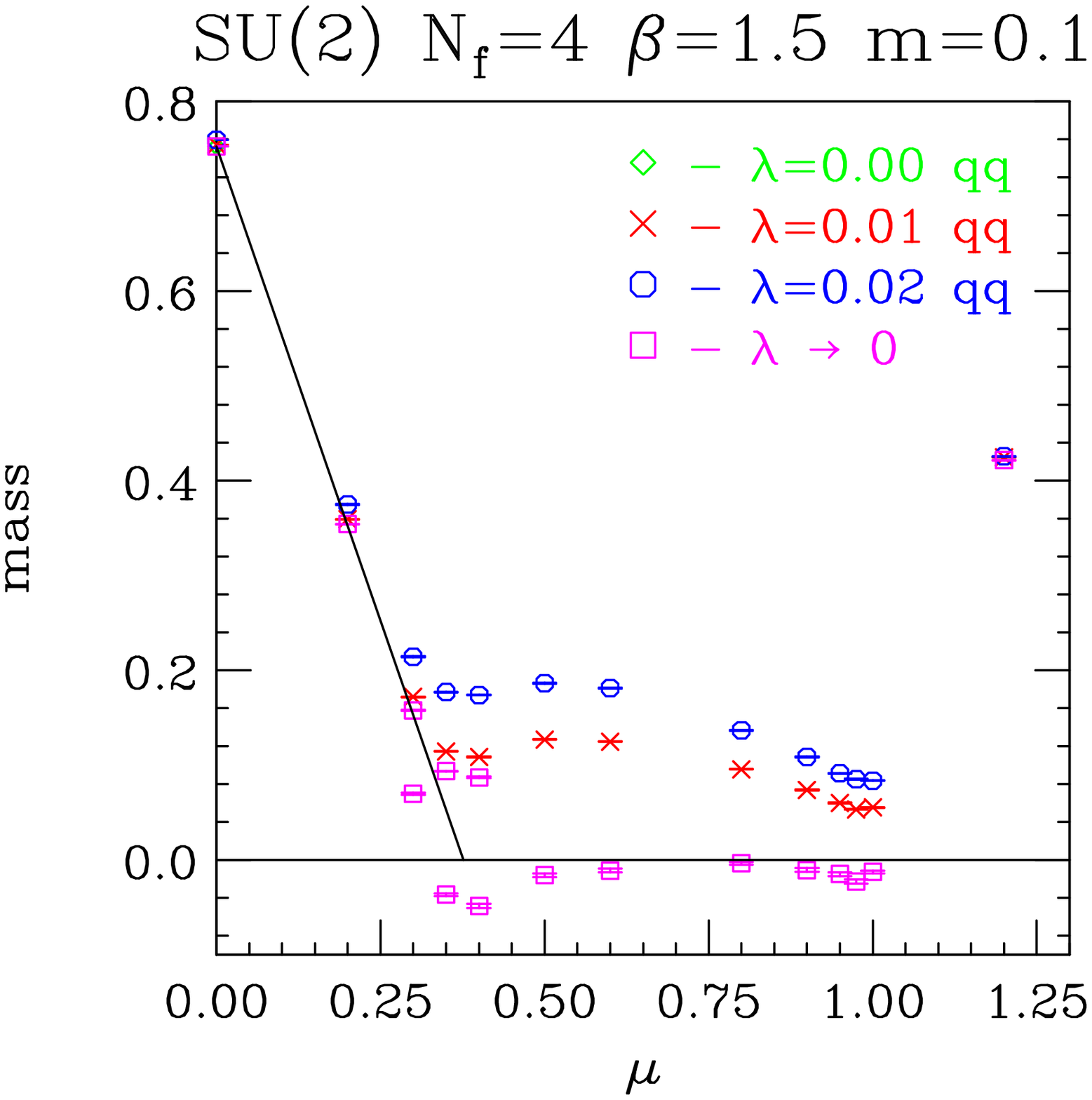}}                                          
\caption{Mass of the diquark state expected to become a Goldstone boson in
the diquark condensed phase, as a function of $\mu$ for $\lambda=0.1$ and
$\lambda=0.2$. The line is the expected linear behaviour expected for the
symmetric phase. Also included are the point at $\lambda=0, \mu=0$ and
$\lambda \rightarrow 0$ extrapolations.}
\label{fig:goldstone}
\end{figure}

In figure~\ref{fig:goldstone} we show the masses from fitting the propagator
for the diquark state produced by applying the operator
$\chi^T\tau_2\chi - \bar{\chi}\tau_2\bar{\chi}^T$ to the vacuum, to the form
\begin{equation}
P_G(T) = A \left\{ \exp[-m_G T] + \exp[-m_G(N_t-T)] \right\},
\end{equation}
valid for large temporal separations $T$. As argued in section~2, this is the
expected Goldstone boson for $\mu > \mu_c$ and $\lambda=0$.
The effective Lagrangian approach indicates that, in the low $\mu$ phase,
this mass should be quadratic in $\lambda$, while in the high $\mu$ phase, it
should vanish as $\sqrt{\lambda}$. We have performed extrapolations based on
this and plotted them in the figure. For the 3 points closest to the transition
$\mu=0.3, 0.35, 0.4$, we have performed both extrapolations.
For $\mu < m_\pi/2$ and $\lambda=0$, we expect
\begin{equation}
m_G = m_\pi - 2\mu
\end{equation}
with $m_\pi = m_\pi(\mu=0)$. We see that for $\mu \le 0.3$, the points obtained
from quadratic extrapolation lie on this straight line, the point at $\mu=0.35$
is close to the line and that for $\mu=0.4$ lies above this line. All points
for $0.35 \le \mu \le 1.0$ obtained by square root extrapolation are negative.
The points at $\mu=0.35,0.4$ show the most significant negative values (although
both are greater than $-0.05$), and should be considered as transitional. The
other points are so close to zero that one can easily attribute the difference
as being due to a combination of the systematic errors in extracting masses
from point source propagators and higher order terms in the extrapolation.
Thus our results are consistent with a transition to a phase with a massless
(Goldstone) scalar diquark at  $\mu \approx m_\pi/2 = 0.37622(5)$, as expected.
Finally the large value of this diquark mass at $\mu = 1.2$ is further
indication that the system has passed through the saturation phase transition.

\begin{figure}[htb]
\epsfxsize=4in
\centerline{\epsffile{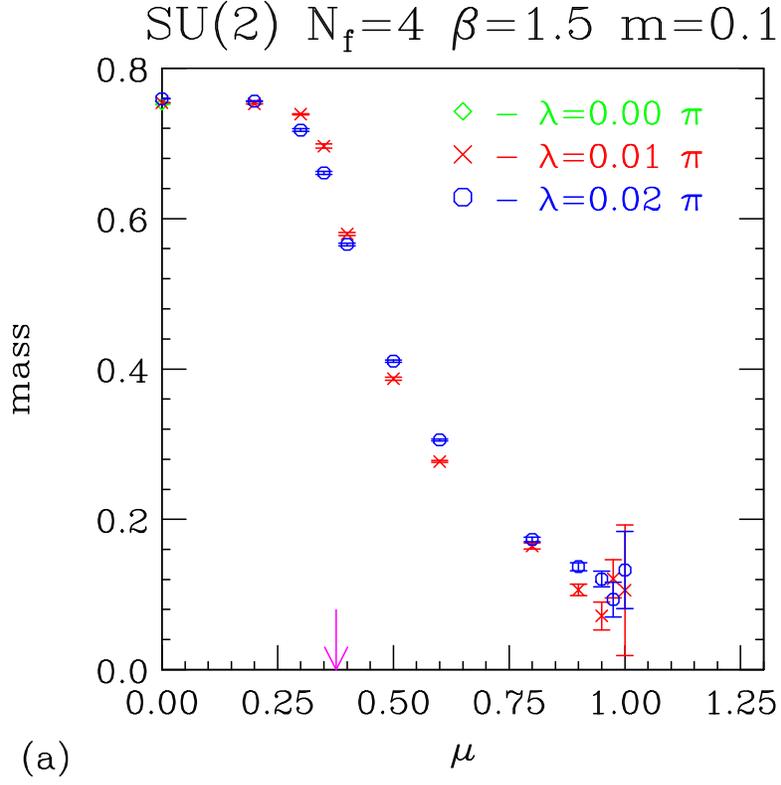}}
\vspace{0.25in}
\centerline{\epsffile{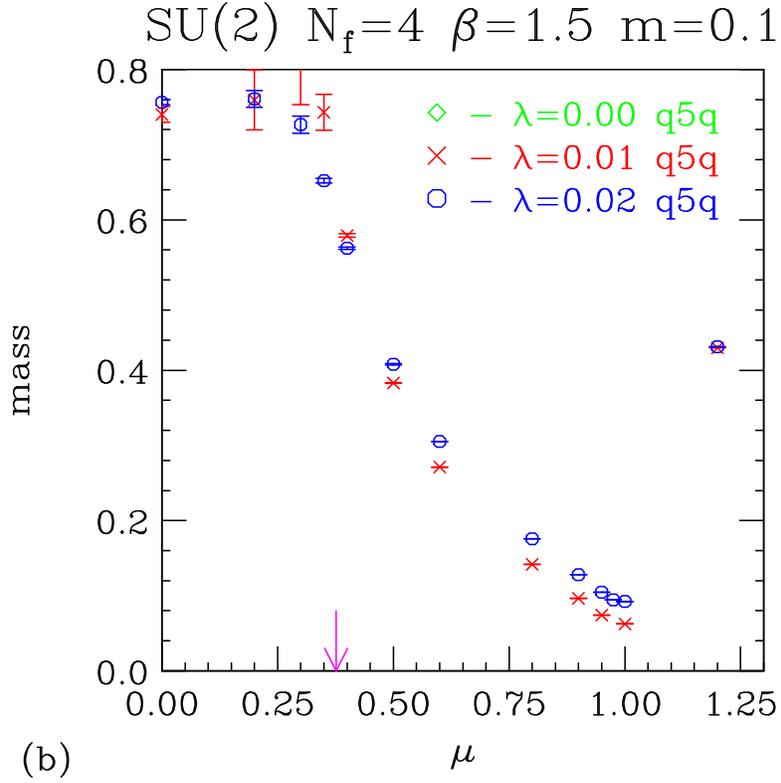}}
\caption{Pion (a) and pseudoscalar diquark (b) masses as functions of $\mu$.}
\label{fig:pseudo}
\end{figure}

We now turn to consideration of the other potential Goldstone bosons of the
theory. Since at this quark mass, $m_\pi^2$ is relatively large, we concentrate
on those states which are expected to have masses of order $m_\pi$ or less over
the range of interest. For this reason, we consider the pion itself. For
$\mu < \mu_c$, its mass should remain constant at $m_\pi$, and it is created
by the operator $\bar{\chi}\epsilon\chi$. For $\mu > \mu_c$, it is expected to
mix with the pseudoscalar diquark state created from the vacuum by the operator
$\chi^T\tau_2\epsilon\chi + \bar{\chi}\tau_2\epsilon\bar{\chi}^T$. Rather than
trying to diagonalize the propagator over these 2 states, we have instead
calculated the diagonal propagators for each of these states separately. These
we fit to the form
\begin{equation}
P_\pi(T) = A \left\{ \exp[-m_\pi T] + \exp[-m_\pi(N_t-T)] \right\}
         + B (-1)^T \left\{ \exp[-m_{b1} T] + \exp[-m_{b1}(N_t-T)] \right\}
\end{equation}
or the form with $B=0$, and a similar form for the diquark state. These masses
are plotted in figure~\ref{fig:pseudo}.
Note that these 2 graphs are almost identical.
For $\mu > \mu_c$, this is expected, since the 2 states mix. For $\mu < \mu_c$
this is not, a priori, expected since the diquark state should exhibit a linear
decrease with $\mu$ for $\lambda=0$. However, since $\lambda \ne 0$ this gives
a small mixing with the pion state, which is then the lowest mass particle 
contributing to this pseudoscalar diquark propagator in this region. It is
precisely because this mixing is so small that the errors are so large. The
final indicator that this is the correct interpretation was that the $\mu=0$,
$\lambda=0$ propagator yielded no sensible mass fit, a sure indicator that the
mass was large where there can be no mixing. The pion mass fits in this low
$\mu$ domain are consistent with the expectation that the $\lambda=0$ pion
mass is independent of $\mu$ in this region. For $\mu > \mu_c$ these masses
fall faster than the expected $m_\pi^2/2\mu$. At least part of the reason for
this more rapid falloff is the saturation effect. Note that the reason that
this mass tends to zero at large $\mu$ (before the saturation
transition) is because the pseudoscalar diquark would be a Goldstone boson if
$m=0$, and as $\mu$ increases the relative importance of $m$ diminishes. The
fact that the $\pi$ mass becomes more poorly determined for large $\mu$ is
because here, the lowest mass state in this channel is predominantly the
pseudoscalar diquark.

It is interesting to contrast our findings with those of a study of adjoint
quarks in 2-colour QCD \cite{hmmoss} with $\lambda$ set to
zero, implying strictly no mixing; in this case $m_\pi$ was found to rise
as $m_\pi\approx2\mu$ for $\mu>\mu_c$, with the signal becoming appreciably
noisier in the high density phase. Both behaviours are in accord with the 
predictions of chiral perturbation theory for a meson formed from quarks 
with a symmetric combination of quantum numbers under the residual global 
symmetry, the difference arising due to the distinct Dyson indices of 
each model \cite{kstvz}.

\section{Conclusions}

Two-colour lattice QCD with one staggered quark species (4 continuum flavours)
has been studied at finite chemical potential $\mu$ for quark number. We have
shown conclusive evidence that this theory undergoes a phase transition to a
phase characterized by a diquark condensate which spontaneously breaks
quark-number, at $\mu=\mu_c \sim m_\pi/2$. This transition appears to be second
order. The simulations were performed at an intermediate value of the coupling
(close to $\beta_c$ for the finite temperature transition at $\mu=0$ on a
lattice with $N_t=4$). Because of this, the relevant symmetry was the lattice
flavour symmetry $U(2)$ rather than the $SU(8)$ of the continuum 4-flavour
theory. We have presented convincing evidence that the scalar diquark is the
Goldstone boson associated with the spontaneous breaking of quark number for
$\mu > \mu_c$. In addition we have measured the lightest mass in the pion
channel as a function of $\mu$. For $\mu \lesssim m_\pi$ the chiral condensate
and diquark condensate are well described as a single condensate of constant
magnitude which rotates from the chiral direction for $\mu < \mu_c$ towards
the diquark direction as $\mu$ is increased above $\mu_c$. Despite the fact
that the quark mass $m=0.1$ was large enough that $m_\pi$ was not small
relative to the 2-colour QCD scale, we saw good qualitative agreement with
previous calculations in terms of effective Lagrangians (of the chiral
perturbation theory form) for $\mu \lesssim m_\pi$. This includes the fact
that the quark number density increases from zero at $\mu = \mu_c$.

As $\mu$ becomes large, this relationship with effective Lagrangians is no
longer valid. Both chiral and diquark condensates decrease towards zero as
$\mu$ becomes large. However, because at least part of this is due to the
fact that $j_0$ saturates at 2 fermions/site, a finite lattice spacing artifact,
and because we see large finite size effects in the condensates for these $\mu$
values, studies at smaller lattice spacings as well as on larger lattices
would be needed to determine how much of this observed high $\mu$ behaviour is
real.

We are currently extending these calculations to lower quark mass ($m=0.025$)
where we should be able to measure the complete spectrum of Goldstone and
pseudo-Goldstone bosons. For this $m$, the pion mass should be half that at
$m=0.1$ and the assumptions of the effective Lagrangian approach should have
more validity, allowing a more quantitative analysis. In addition, there is
a larger range of $\mu/m_\pi$ between $\mu_c$ and the turnover point. This
means we can hope to measure the critical index $\beta_{qq}$ for this 
transition. (Preliminary attempts to extract this index from short runs at
relatively strong coupling were reported in our letter on finite $T$ and $\mu$.)

Let us now compare 2-colour QCD with 3-colour QCD, at finite $\mu$. Most of
this discussion is condensed from the published literature \cite{arw,rssv,as}.
Since for 2-colour QCD, the diquark condensate is a colour singlet, the
spontaneous breaking of quark number is realized in the Goldstone mode. This
contrasts with true (3-colour) QCD where the condensate is, of necessity,
coloured and the symmetry breaking is realized in the Higgs mode. To see
similarities, we consider the mode of symmetry-breaking for normal QCD with 2
light quark flavours. Here the expected condensate is a flavour singlet and a
colour anti-triplet, i.e. antisymmetric in both flavour and colour. The
pattern of colour breaking is $SU(3) \times U(1)_q \rightarrow SU(2) \times
U(1)_Q$ where $q$ and $Q$ refer to quark number before and after the breaking.
The 5 gluons corresponding to symmetries broken by the condensate,
gain masses via the Higgs mechanism by combining with the 5 would-be Goldstone
bosons associated with colour/quark-number breaking. With respect to the
remnant $SU(2)$ colour symmetry, the condensate is a flavour {\it and} colour
singlet. This is precisely the condensate that would be formed from the 2
quarks which are in a colour doublet with respect to this unbroken $SU(2)$
(the third quark is a singlet with respect to this group) if we ignored the
interactions of the gluons which gain masses due to the Higgs mechanism. Since
the gluon masses produced from the Goldstone bosons produced in this manner
are of order $\alpha_s^{1/2}(\Lambda_{QCD})f_\pi$, ignoring such light particle
interactions is presumably a rather drastic approximation. However, one might
hope that it has at least qualitative validity. (Note that $f_\pi$ is of the
order of $100$~MeV, $\alpha_s(\Lambda_{QCD}) \sim 1$, giving gluon masses of
the same order of magnitude as previous estimates.) Since this remnant $SU(2)$
theory is also at finite $\mu$ this condensate spontaneously breaks only 1
symmetry, quark number. As seen above, the Goldstone boson associated with
this breaking gives mass to one of the gluons via the Higgs mechanism.

Finally, we note that with an appropriate reinterpretation of the fields,
2-colour QCD with a finite chemical potential for quark number can be 
reinterpreted as 2-colour QCD with a finite chemical potential for isospin.
It is clear that one can simulate true 3-colour QCD for 2 flavours with a
finite chemical potential for isospin. Such a program is underway. The
2-colour theory can be used as a guide to the pattern of symmetry breaking and
what to expect.

\section*{Acknowledgements}

This work was partially supported by the NSF under grant NSF-PHY96-05199 and
by the U.~S. Department of Energy under contract W-31-109-ENG-38. 
SJH and SEM were supported by EU-TMR contract no. ERBFMRX-CT97-0122. These
simulations were performed on the IBM SP and Cray SV1's at NERSC and on the
IBM SP and Cray T90 at NPACI. DKS would line to thank C.~K.~Zachos for useful
discussions.

\end{document}